# CONCEPT-BASED INDEXING IN TEXT INFORMATION RETRIEVAL

## Fatiha Boubekeur[1] and Wassila Azzoug[2]

[1]Department of Computer Science, Mouloud Mammeri University of Tizi-Ouzou, Algeria
amirouchefatiha@mail.ummto.dz
[2] Limose Laboratory, Department of Computer Science, M'Hamed Bouguera, University of Boumerdes, Algeria
azzoug_w@umbb.dz

## Abstract

*Traditional information retrieval systems rely on keywords to index documents and queries. In such systems, documents are retrieved based on the number of shared keywords with the query. This lexical-focused retrieval leads to inaccurate and incomplete results when different keywords are used to describe the documents and queries. Semantic-focused retrieval approaches attempt to overcome this problem by relying on concepts rather than on keywords to indexing and retrieval. The goal is to retrieve documents that are semantically relevant to a given user query. This paper addresses this issue by proposing a solution at the indexing level. More precisely, we propose a novel approach for semantic indexing based on concepts identified from a linguistic resource. In particular, our approach relies on the joint use of WordNet and WordNetDomains lexical databases for concept identification. Furthermore, we propose a semantic-based concept weighting scheme that relies on a novel definition of concept centrality. The resulting system is evaluated on the TIME test collection. Experimental results show the effectiveness of our proposition over traditional IR approaches.*

## 1. INTRODUCTION

Information retrieval (IR) is concerned with selecting from a collection of documents, those that are likely to be relevant to a user's information need expressed using a query. Three basic functions are carried out in an information retrieval system (IRS): document and information need representation, and matching of these representations. Document representation is usually called indexing. The main objective of document indexing is to associate a document with a descriptor represented by a set of features manually assigned or automatically derived from its content. Representing the user's information need involves a one step or multi-step query formulation by means of prior terms expressed by the user and/or additive information driven by iterative query improvements like relevance feedback [1]. The main goal of document-query matching, also called query evaluation, is to estimate the relevance of a document to the given query. Most of IR models handle during this step an approximate matching process using the frequency distribution of query terms over the documents to compute the relevance score. This score is used as a criterion to rank the list of documents returned to the user in response to his query.

Traditional IRS are based on the well known technique of "bag of words" (BOW) representation expressing the fact that both documents and queries are represented as bags of lexical entities, namely keywords. A keyword may be a simple word (as in "*computer*" ) or a compound word (as in "*computer science*").  Weights are associated with document or query keywords [2], [3] to express their importance  in the considered material.  The weighting scheme is generally based on variations of the well known *tf\*idf* formula [4].

A key characteristic of such systems is that  the degree of document-query matching depends on the number of shared keywords. This leads to a "*lexical focused*" relevance estimation which is less effective than a "*semantic  focused*" one [5]. Indeed, in such IRS, relevant documents are not retrieved if they do not share words with the query, and irrelevant documents that have common words with the query are retrieved even if these words have not the same meaning in the document and the query. The problems mainly stem from the richness in terms of expressive power, yet the *synonymy* and *polysemy* inherent in natural language.

To address this shortcoming, in this paper, we propose a novel approach to indexing and weighting documents and queries using semantic entities, the concepts, in addition to lexical entities, the keywords. In our concept-based approach, concepts are identified from the content of the document (or the query), and weighted according to both their frequency distribution and their semantic similarity (relatedness) to other concepts in the document (or the query). To identify accurate concepts in the considered material, the approach relies on a two-steps word sense disambiguation (WSD) approach based on the joint use of WordNet [6] and its extension WordNetDomains [7].

The remainder of the paper is structured as follows: Section 2 introduces an essential background about WordNet, WordNetDomains and WordNet-based semantic similarity measures. Section 3 discusses related work in the area of concept-based indexing and situates our contribution. Section 4 details our approach. Experimental results are presented in section 5.  Section 6 concludes the paper.

## 2. BACKGROUND

### 2.1. WordNet

WordNet [6] is an electronic lexical database which covers the majority of nouns, verbs, adjectives and adverbs of the English language, which it structured in a network of nodes and links.

A node also called *synset*, is a set of synonymous terms that are interchangeable in a context. A synset represents a concept or a word sense. A synset is lexically represented by a term from its synonyms. Almost each WordNet synset has a *gloss* expressed in English that defines that synset. A synset's gloss may also contain comments and/or one or more examples of how the words in the synset are used [8]. Table 1 presents the synsets associated with the word "*bank*", among which the first synset, {*depository financial institution, bank, banking concern, banking company*}, is defined by the gloss: -- (*a financial institution that accepts deposits and channels the money into lending activities*"; "*he cashed a check at the bank*"; "*that bank holds the mortgage on my home*").

A link represents a semantic relation between two synsets.  The majority of WordNet's relations connect terms from the same part of speech (POS).  Thus, WordNet really consists of four sub-nets, one each for nouns, verbs, adjectives and adverbs, with few cross-POS pointers[1]. For example, the main encoded relations among noun synsets are the following:

- the subsumption relation or *is-a* relation (also called hypernymy/hyponymy) associates a general concept (the hypernym) to a more specific one (its hyponym). For example, the noun *bank#n#1*[2] has as hyponyms *credit union*, *Federal Reserve Bank*, *agent bank*, *commercial bank*, *state bank*, etc. The *is-a* relation thus organizes WordNet synsets into a hierarchy.
- the part-whole relation (or Meronymy/holonymy), associates a concept *Y* (holonym) to its part *X* (meronym). For example, *building* is a holonym of *window*. And conversely *window* is a meronym of *building*.

Table 1.  WordNet synsets of the word "bank"

| The noun bank has 10 senses (first 9 from tagged texts) |
| --- |
| 1. (883) depository financial institution, bank, banking concern, banking company -- (a financial institution that accepts deposits and channels the money into lending activities; "he cashed a check at the bank"; "that bank holds the mortgage on my home")<br>2. (99) bank -- (sloping land (especially the slope beside a body of water); "they pulled the canoe up on the bank"; "he sat on the bank of the river and watched the currents")<br>3. (76) bank -- (a supply or stock held in reserve for future use (especially in emergencies))<br>4. (54) bank, bank building -- (a building in which the business of banking transacted; "the bank is on the corner of Nassau and Witherspoon")<br>5. (7) bank -- (an arrangement of similar objects in a row or in tiers; "he operated a bank of switches")<br>6. (6) savings bank, coin bank, money box, bank -- (a container (usually with a slot in the top) for keeping money at home; "the coin bank was empty")<br>7. (3) bank -- (a long ridge or pile; "a huge bank of earth")<br>8. (1) bank -- (the funds held by a gambling house or the dealer in some gambling games; "he tried to break the bank at Monte Carlo")<br>9. (1) bank, cant, camber -- (a slope in the turn of a road or track; the outside is higher than the inside in order to reduce the effects of centrifugal force)<br>10. bank -- (a flight maneuver; aircraft tips laterally about its longitudinal axis (especially in turning); "the   plane went into a steep bank"). |

## 2.2. WordNet-Based Semantic Relatedness Measures

Numerous approaches to measuring semantic relatedness between WordNet synsets are proposed in the literature, which are classified into two main types: path-based measures and information content-based measures.

- In path-based measures, WordNet is viewed as a graph of concepts and semantic relatedness between two concepts is measured through edge-counting (ie. path length) between their corresponding nodes in the graph.  The underlying principle is that the shortest  path from one node to another is, the more similar the concepts are. Leackock and Chodrow measure [9] and Wu-Palmer measure [10] range from this category.

---

[1] http://wordnet.princeton.edu/

[2] *bank#n#1* refers to the first sense of the noun *bank*  in WordNet.

- − In information content-based measures, semantic relatedness between two concepts is measured through the information they share in common. Resnik measure [11], lin measure [12], and Jiang and Conrath measure [13] range from this category.

We refer the interested reader to [14] for an overview on the cited measures.

### 2.3. WordNetDomains

WordNetDomains [7] is an extension of WordNet lexical database that results from the annotation of each synset with one or more domain label from a set of 176 domains hierarchically organized through the subsumption (specialization/generalization) relation (for example, *Tennis* is a more specific domain than *Sport*, and *Architecture* is a more general domain than *Buildings*).

Part of the WordNetDomains hierarchy is given in Table 3. The *Top-Level* domain is the root of this hierarchy. *Factotum* is a functional domain (as opposed to semantic one) that includes generic synsets which are hard to classify in any particular domain, and and Stop senses synsets (such as colors, numbers, etc.) which appeared frequently in different contexts [15]. Factotum is independent from the *Top-Level* domain and its hierarchy.

Table 2. Domains associated with the synsets of the word « *bank* »

| WordNet Synsets | Associated Domains |
|---|---|
| 1. (883) depository financial institution, bank, banking concern, banking company -- (a financial institution that accepts deposits and channels the money into lending activities; "he cashed a check at the bank"; "that bank holds the mortgage on my home") | ECONOMY, |
| 2. (99) bank -- (sloping land (especially the slope beside a body of water); "they pulled the canoe up on the bank"; "he sat on the bank of the river and watched the currents") | GEOGRAPHY, GEOLOGY |
| 3. (76) bank -- (a supply or stock held in reserve for future use (especially in emergencies)) | ECONOMY |
| 4. (54) bank, bank building -- (a building in which the business of banking transacted; "the bank is on the corner of Nassau and Witherspoon") | FACTOTUM, ECONOMY |
| 5. (7) bank -- (an arrangement of similar objects in a row or in tiers; "he operated a bank of switches") | FACTOTUM |
| 6. (6) savings bank, coin bank, money box, bank -- (a container (usually with a slot in the top) for keeping money at home; "the coin bank was empty") | ECONOMY |
| 7. (3) bank -- (a long ridge or pile; "a huge bank of earth") | GEOGRAPHY, GEOLOGY |
| 8. (1) bank -- (the funds held by a gambling house or the dealer in some gambling games; "he tried to break the bank at Monte Carlo") | ECONOMY, PLAY |
| 9. (1) bank, cant, camber -- (a slope in the turn of a road or track; the outside is higher than the inside in order to reduce the effects of centrifugal force) | ARCHITECTURE |
| 10. bank -- (a flight maneuver; aircraft tips laterally about its longitudinal axis (especially in turning); "the plane went into a steep bank") | TRANSPORT |

## 3. RELATED WORK

Concept-based indexing represents both documents and queries using semantic entities, the concepts, instead of (or in addition to) lexical entities, the keywords. Retrieval is then performed in this conceptual space. Concept-based indexing approaches hold the promise that representing documents and queries (or enhancing their BOW representation) using concepts will result in a retrieval model that is less dependent on the index terms [16]. Indeed, in such a model, documents could be retrieved even when the same concept is described by different terms in the query and the documents, thus alleviating the *synonymy* problem and increasing recall[3] [17]. Similarly, if the correct concepts are chosen for ambiguous words appearing in the query and in the documents, non-relevant documents would not be retrieved, thus alleviating the *polysemy* problem and increasing precision[3].

Table 3. Part of WordNetDomains hierarchy.

| | | | | |
|---|---|---|---|---|
| Top_level | Humanities | History | | |
| | | Linguistics | Grammar | |
| | | Literature | Philology | |
| | | Philosophy | Psychoanalysis | |
| | | Art | Music | |
| | | | Plastic_Arts | Jewellery |
| | | | | Sculpture |
| | | | Theatre | |
| | | | Cinema | |
| | | Paranormal | | |
| | | … | | |
| | … | | | |
| | Pure_Science | Biology | Anatomy | |
| | | Animals | | |
| | | Earth | Geology | |
| | | | Geography | |
| | | Mathematics | | |
| | | Physics | Acoustics | |
| | | … | | |
| | Social_Science | Economy | Finance | Money |
| | | Politics | | |
| | | Fashion | | |
| | | Military | | |
| | | … | | |
| Factotum | Quality | | | |
| | Number | | | |
| | … | | | |

Concept-based indexing relies on concepts identified from the content of the document and the queries based on linguistic knowledge resources (such as dictionaries, thesauri, ontologies, etc.). These concepts describe the content of a given text to different extents [18]. To capture this characteristic, each concept is assigned a weight that reflects its relative importance in the indexed text. The indexing process thus runs in two main steps: (1) concept identification and (2) concept weighting.

---

[3] Recall and precision are two measures used to estimate the effectiveness of an IRS respectively in terms of the ratio of relevant documents that are retrieved, and the ratio of retrieved documents that are relevant.

### 3.1. Concept Identification

Concept identification aims at assigning documents terms[4] to the corresponding entries in the ontology (or any other linguistic resource). For this aim, representative keywords are first identified in each document, using classical indexing techniques (tokenization, lemmatization, stop words elimination, etc.). More complex processes can also be integrated to recognize multiword features (nominal phrases, collocations ...) [19]. These terms are then mapped onto the ontology in order to identify the corresponding concepts (or senses). An ambiguous (polysemic) term may correspond to several entries (senses) in the ontology, it must be disambiguated. To disambiguate a term, WSD approaches generally exploit local context and definitions from the ontology [20], [21], [22], [23], [24]. The underlying idea is to estimate the "semantic relatedness" between each sense associated with the target term and the other senses from its local context. Formally, the disambiguation process relies on the computation of a score for each concept on the basis of its semantic relatedness to other concepts in the document context. The concept which maximizes the score is then retained as the correct sense of the target term in the document. Approaches in [22], [23], [25], [26], [27] are based on these principles.

To disambiguate an ambiguous word, Voorhees [22], classified each synset of this word based on the number of words collocated between a *neighborhood* of this synset in the WordNet *is-a* hierarchy and the local context (the sentence in which the word occurs) of the target word. The best classified synset is then considered as the appropriate sense of the ambiguous word.

In a similar approach, Katz et al [26] defined the local context of a word as the ordered list of words starting from the closest useful word to the left or right neighborhood until the target word. To disambiguate a word, Katz et al. first extract all non-empty words (called *selectors*) from the local context of the target word. The set $S$ of selectors is then compared to WordNet synsets. The synset that shares a maximum number of words with $S$ is selected as the appropriate sense of the target word.

To disambiguate an ambiguous word, Khan et al. [27], proposed an approach based on the semantic closeness of concepts. The semantic closeness of two concepts is calculated by a score based on their mutual minimal distance in a given domain-oriented ontology. The concepts that have the highest score are then selected.

Based on the principle that among the various possible senses (WordNet synsets) of a word, the most appropriate one maximizes its relations with other possible senses in the document, Baziz et al. [23], assigned a score to each sense of each ambiguous word. The score of a given sense is obtained by summing the values of its semantic relatedness to other possible senses in the document. The sense having the highest score is then selected as the appropriate sense of the associated word.

In our approach proposed in [25], the score associated with a possible sense of a word is based on its semantic similarities with other candidate concepts (senses) in the document, balanced by the occurrence frequencies of the considered features.

In a more recent work, the authors in [28] proposed a domain-oriented disambiguation approach that relies on first identifying the correct domain of a word in the document based on WordNetDomains, and then disambiguating the word in the identified domain based on WordNet.

---

[4] "*terms*" refers to simple words (keywords) or multi-words (collocations, noun phrases, etc.)

The correct domain of a word is selected on the basis of its frequency distribution in the context of the target word.

Based on a similar principle, we proposed in [29] to disambiguate a domain on the basis of its semantic relatedness to other domains associated with other terms in the same context. Disambiguating a word in its selected domain is based on a score related to the intensity of its semantic relatedness to other words from its context.

## 3.2. Concept Weighting

The concepts weighting aims at evaluating the importance of each concept in a document's (or query) content. This importance can be estimated either statistically through its frequency distribution in the document (or query) content using a normalized version of the classical *tf*idf* scheme as in [22], [23], [30], or semantically through its centrality (ie. its semantic relatedness to other concepts) in the document (or query) as in [18], [29], [31], [32].

In statistical weighting approaches, concepts are considered through the terms which represent them. Hence, concepts weighting consists on terms weighting. The weighting approaches of Harrathi et al. [30], Baziz et al. [23] and Voorhees [22] rely on this principle. Based on the extended vector space model introduced in [9], in which every vector consists of a set of sub-vectors of various concept types (called *ctypes*), Voorhees [22] proposed to weight the concepts using a normalized classic *tf*idf* scheme. The approach proposed by Baziz et al. [23], extends the *tf*idf* scheme to taking into account the compound terms (or multi-words). Indeed, in this so-called *Cf*idf* approach, the weight of a compound term is based on the cumulative frequency of the term itself and of its components. In [30], the proposed *tf*ief* weighting scheme is an adapted version of *tf*idf* to concepts weighting relatively to a given element of an XML document.

In semantic weighting approaches, concepts are considered through the senses they represent. Hence, concepts weighting aims at evaluating the importance of the corresponding senses in the document's (or query) content. In [31], [32], this importance is estimated through the number of semantic relations between the target concept and other concepts in a document. These relations are also weighted in [18]. In [31], the number of relations a concept has with other concepts in the document defines its *centrality*. The authors in [32] combine *centrality* and *specificity* to estimate the importance of a concept in a document. Concept specificity represents its depth in the WordNet hierarchy. In our work presented in [25], we proposed to weight compound terms (representing concepts) on the basis of their semantic relatedness to their corresponding *sub-terms* (components) and *sur-terms* (containers). Practically, the weight of a term is estimated through a probabilistic measurement of the relatedness of all its possible senses to other senses associated with its sub-terms and sur-terms taking into account their respective frequencies in the document. In our semantic indexing approach proposed in [24], the importance of a concept in a document is expressed through its (cumulative) semantic relatedness to other concepts in the document. This has been combined with the statistical frequency measure in our proposal in [29].

In the present paper, we propose a novel concept-based indexing approach that relies on the joint use of WordNet and WordNetDomains for extracting representative concepts from documents and queries, and assigning them semantic weights that reflect their importance in the indexed materials.

## 4. CONCEPT-BASED DOCUMENT INDEXING

### 4.1. Definitions and Notations

Let $m$ be a word of the text to be indexed.

— We call *instance* of $m$, each occurrence $m_i$ of $m$ in the given text.

— An instance of word is a word. It is associated with a single part of speech (*noun, verb, adverb, etc.*) in the sentence where it appears.

— An instance $m_i$ of word $m$ appears in a sentence. The set of index terms of this sentence defines the *local context* of $m_i$ which is noted $\zeta_{L_i}$.

— The *global context* of $m_i$ is the union of all local contexts in which $m_i$ appears with the same part of speech. Each global context thus defines a different meaning for $m$. The global context of $m_i$ is noted by $\zeta_i$.

— The *local expression* of $m_i$ of size $s+1$ is the character string obtained from concatenating the word $m_i$ and the $s$ successive words located immediately to the right of $m_i$ by using an underscore (_) between them.

— The size of a local expression is the number of words it contains.

### 4.2. Description of the Approach

The key objective of our approach is to represent the document (respectively the query) by a semantic index composed of two types of terms: concepts and orphan keywords.

— Concepts are (unambiguous) WordNet entries (synsets) identified from the text of the document. The concepts are denoted by either simple words or collocations.

— Orphan keywords are (non-empty) simple words of the document that do not have entries in WordNet.

Concepts are first identified in the document (respectively in the query) by means of an identification/disambiguation process (which also allows to identifying orphan keywords) and then weighted. Thus, the indexing process mainly runs in two steps: first step is concept identification, and second step is the concept weighting.

### 4.2.1 Concept Identification

The main objective of this step is to identify the representative concepts of a given document (respectively query). Concepts are entries in WordNet which correspond to the representative terms of the indexed document (respectively query). Concepts identification implies (a) identifying index terms from the considered text and then mapping them to WordNet synsets, and (b) disambiguating ambiguous terms (an ambiguous term corresponds to more than one synset).

#### 4.2.1.1 Identifying Index Terms

The purpose of this step is to identify the set of representative terms (words or collocations) of the considered material relying on WordNet. This step begins by the identification of collocations. For this aim, we first built a list $\varphi_{Coloc}$ of all WordNet's collocations. Then, for each analyzed word's instance $m_i$, we extract from $\varphi_{Coloc}$ the set $\varsigma_i$ of collocations that begin with $m_i$. $\varsigma_i$ is first ranked by decreasing order of its elements size, then each element in $\varsigma_i$ is projected on the $m_i$'s local expression $E_i$ of size $|\varsigma_i|$. If a collocation matches with a local expression, it is retained and inserted into the set $\xi_{Expres}$ of identified collocations. If no collocation matches with the local expression of $m_i$, $m_i$ is a simple word. If this simple word has an entry in WordNet, it will be inserted into the set $\xi_{Simples}$ of simple words. Otherwise it will be added to the orphans set $\xi_{Orphel}$ . The terms identification algorithm is given in Table 4.

### 4.2.1.2 Term Disambiguation

This step aims at assigning a meaning to each ambiguous index term based on the context in which it occurs. Collocations are almost unambiguous terms, hence the disambiguation process which mainly relies on WordNet will only involve the simple words which have entries in this lexical database. Therefore only the words of $\xi_{Simples}$ are concerned.

A word of $\xi_{Simples}$ can have several entries (synsets) in WordNet which correspond to different senses. The purpose of this step is to select the appropriate synset of the word based on its context.To disambiguate words, we propose an approach based on three disambiguation levels:

– The first level is part of speech (POS) identification. This level aims at determining the POS of each word $m_i$ in the document using the Stanford POS Tagger.
– The second level is domain disambiguation. This level aims at identifying the usage domain of a word in the context of a document (or query). Domain identification relies on the use of WordNetDomains. This disambiguation level will limit the number of senses to be discussed in the third level.
– The third level is word sense disambiguation. It aims at selecting among the possible senses of the word in the selected domain the most appropriate sense in the document.

Table 4. Index terms identification algorithm

---

**Input** : document $d$.

**Output** : $\xi_{Expres}$ , $\xi_{Simples}$ , $\xi_{Orphel}$

**Procedure**: Let $m_i$ be the next word to be analyzed in $d$.

    Begin

        1. Compute $\varsigma_i = \{C^i_1, C^i_2, \dots C^i_n\}$ the set of collocations beginning with $m_i$

        2. Order $\varsigma_i$ as : $\varsigma_i = \{C^i_{(1)}, C^i_{(2)},\dots, C^i_{(n)}\}$ where $(j)_{1..n}$ is an index permutation such that $|C^i_{(1)}| \geq |C^i_{(2)}| \geq \dots \geq |C^i_{(n)}|$, where$|C^i_{(j)}|$ is the size of collocation $C^i_{(j)}$

        3. Bool<- false ;

        4. While bool=false do:

        5. $C^i_{(j)}$ := next collocation in $\varsigma_i$

        6. Compute the local expression $E_i$ of size $|C^i_{(j)}|$

        7. if $E_i = C^i_{(j)}$ then bool <- true ; enddo;

        8. if (bool) then insert $E_i$ into $\xi_{Expres}$

           else if $m_i$ is a non empty word then

              begin

                if $m_i$ has an entry in WordNet then insert $m_i$ into $\xi_{Simples}$

                else insert $m_i$ into $\xi_{Orphel}$ ;

              end ;

    End.

---

**4.2.1.2.1 Part of speech Identification**

A word may have several instances in a document, each of which is characterized by its POS. The objective of this step is to identify the POS of each instance in a given material. For this aim, we simply rely on utilizing the Stanford POS Tagger. As the synsets associated with a word's instance are grouped in WordNet according to their part of speech, this step aims to limit the synsets to be examined in the next disambiguation steps to those having the same POS as the target instance.

**4.2.1.2.2. Domain Disambiguation**

WordNet synsets are labeled in WordNetDomains by means of domain labels. An ambiguous word has several associated senses (synsets) in WordNet each of which can belong to one or more domains in WordNetDomains. The aim of this step is to select the correct domain of a word in the context of the document.

Assuming that the appropriate domain of a word $m_i$ is likely to be highly related to the other domains of its (global) context $\zeta_i$, we assign each domain $D_j$ associated with each sense (synset) of $m_i$ a score based on its semantic relatedness with other domains $D_k$ associated with the other terms $t_k$ ( $t_k \in \{ \xi_{Simples} \cup \xi_{Expres} \}$ )belonging to $\zeta_i$ .The domain that maximizes this score is then selected as the appropriate domain of the word $m_i$ in the document. Formally:

$$Score\left(D_j\right) = \arg\max_j \left( \sum_{t_k \in \zeta_i} \sum_{k \in [1..n]} Sim\left(D_j, D_k\right) \right)$$

where $Sim\left(D_j, D_k\right)$ denotes the semantic relatedness of domains $D_j$ and $D_k$ estimated through the Wu-Palmer [10] measure which we adapt to the WordNetDomains hierarchy as follows:

$$Sim\left(D_j, D_k\right) = \frac{2 * depth\ (D^*)}{depth\ (D_j) + depth\ (D_k)}.$$

Where:

- $D^*$ is the least common subsumer of $D_j$ and $D_k$ in the *WordNetDomains* hierarchy.

- $depth\left(D\right)$ is the depth of $D$ in the *WordNetDomains* hierarchy.

Remark._ We work on the assumption that a domain is directly related to other domains in the WordNetDomains hierarchy. Therefore, it makes no sense to use the domain factotum for this disambiguation technique.

**4.2.1.2.3 Word sense disambiguation**

At this stage, every word $m_i$ in $\xi_{Simples}$ is associated with a single domain $D_i$ in its context. But it still can be associated with more than one synset (sense) in this domain. In this case, it must be disambiguated. The aim is to select among all the synsets associated with $m_i$ in $D_i$, the appropriate sense (meaning) of $m_i$ in its context.

Let $S_{i(j)}$ be the set of all synsets associated with the word $m_i$ in the domain $D_j$, and $S_{i(j)}[k]$ the k[th] element of $S_{i(j)}$. To disambiguate the word $m_i$ in its domain $D_j$, we associate a score with each synset $S_{i(j)}[k]$ based on its semantic relatedness with other synsets associated with the other terms of the context. The synset with the highest score is selected as the appropriate meaning of the word $m_i$ in its context. Formally:

$$S_{i(j)}[k] = Arg \max_k \left( \sum_{l \mid t_l \in \zeta_i, \, l \neq i} \sum_{1 \leq n \leq |S_{i(j)}|} Sim\left(S_{i(j)}[k], S_{l(m)}[n]\right) \right)$$

Where $Sim\left(S_{i(j)}[k], S_{l(m)}[n]\right)$ estimates the semantic relatedness (or semantic similarity) between the concepts $S_{i(j)}[k]$ and $S_{l(m)}[n]$ on the basis of the Resnik measure [11] (or any other WordNet-based similarity measure [12], [33] …)

## 4.2.2. Concepts Weighting

The objective of this step is to assign weights to the identified concepts (synsets) which express their importance in the document.

Starting from the idea that the more a concept is locally central in the document and globally central in the collection the more it is representative of the content of the document, we will weight concepts according to their local and global centralities defined as follows:

The *local centrality* of a concept $C_i$ in a document $d$, denoted $cc(C_i, d)$, is defined on the basis of its relevance in the document on the one hand and on its occurrence frequency on the other hand. The concept's relevance is measured through its semantic relatedness with the other concepts in the document. Its frequency is the cumulative frequency of all its representative terms in the document. Formally:

$$cc(C_i, d) = \alpha * tf(C_i, d) + (1 - \alpha) \sum_{i \neq l} Sim(C_i, C_l)$$

Where:

- is a weighting factor used to balance the frequency with respect to the relevance. This factor is determined experimentally,

- $Sim(C^i, C^l)$ measures the semantic similarity between concepts $C^i$ and $C^l$, calculated on the basis of the Resnik measure [11],

- $tf(C^i, d)$ is the occurrence frequency of the concept $C^i$ in the document.

Definition._ A concept $C_i$ is (locally) central in a document $d$ if its local centrality in $d$ is greater than a fixed threshold $s$ (ie. $cc(C_i, d) > s$ ).

The *global centrality* of a concept $C_i$ defines its discrimination power in the collection of documents (that is its ability to discriminate between those documents that contain informative concepts and those that contain non-informative ones). The idea is that a concept that is central in too many documents is non-informative. On the other hand, a concept that is central in few documents is considered more informative. The ratio of documents in which the concept $C_i$ is central defines the *document centrality* of $C_i$, noted $dc(C_i)$. Formally:

$$dc(C_i) = \frac{n}{N}$$

Where $N$ is the total number of documents in the collection, and $n$ is the number of documents in which $C_i$ is a central concept.

The discrimination power (ie. the *global centrality*) of a concept is then seen as a measure of its *inverted document centrality* (denoted by *idc*). Formally:

$$idc(C_i) = \frac{1}{dc(C_i)}$$

The weight associated with a concept $C_i$ in a document $d$ is then defined as a combination of its local centrality and its global centrality. Formally:

$$W(C_{i,d}) = cc(C_i, d) * idc(C_i)$$

This proposed scheme, called *cc-idc*, allows weighting the concepts as well as the orphan keywords. In this latter case, only the *tf* factor of local centrality is considered (WordNet-based semantic similarity doesn't apply for orphan keywords, it is then set to zero).

# 5. EXPERIMENTAL EVALUATION

In this section, we present our experimental evaluation of the proposed approach. Our objective is twofold: (1) measure the effectiveness of our concept-based indexing approach over classical indexing approaches, and (2) study the performance of our semantic weighting scheme (*cc-idc*) over classical weighting schemes.

In the following, we first introduce the experimental setting (test collection and evaluation protocol), then we present and discuss evaluation results.

## 5.1. Experimental Setting

### 5.1.1 Test Collection

For our experiments, we used the Time[5] test collection. This dataset contains 423 documents consisting of newspaper articles from the TIME magazine and a large number of queries (83). Human relevant judgements are associated with each query. We chose to use the 15 queries that provide the most significant results in classical search based on *tf\*idf* weighting.

### 5.1.2. Evaluation Protocol

The approach is implemented through a concept-based vector IR system. In this system, documents and queries are seen as vectors of weighted concepts, and compared through the classical cosine measure. Evaluation is performed according to TREC protocol. It is based on two measures: precision at cutoff $x$, $P @ x$ ($x$ = 5, 10, 20, 50, 100, 200, 500), and mean average precision ($MAP$).

---

[5] ftp://ftp.cs.cornell.edu/pub/smart/time/

Precision at cutoff $x$, $P@x$ is the ratio of relevant documents among the top $x$ returned documents. Formally, if $RR_x$ denote the number of relevant documents among the $x$ first retrieved documents, then:

$$P @ x = \frac{RR_x}{x}$$

$MAP$ is the arithmetic mean over all queries of the average precision (calculated at the ranks in which relevant documents are retrieved) calculated for each query. Formally:

$$MAP = \frac{1}{N} \sum_{j=1}^{N} \left( \frac{1}{Q_j} \sum_{i=1}^{Q_j} P(doc_i) \right)$$

Where:

- $Q_j$ is the number of relevant documents for query $j$,

- $N$ is the total number of queries,

- $P(doc_i)$ is the precision at the rank of the i[th] observed relevant document.

Practically, for each query, the top 100 documents retrieved by the system are examined, and precisions P @ x (x = 1, 2, 3, 4, 5, 10, 15, 20, 30, 40, 50, 100) and MAP are calculated. The *Trec_eval*[6] program is used for these calculations. We then compare the results from our semantic index to those returned by reference systems (or baselines). In our experiments, we consider two baselines:

- The first (denoted *Classic-TF_IDF*) corresponds to a classical index based on *tf*iidf* - weighted keywords,

- The second (denoted *Classic-BM25*) is a classical index based on Okapi-*BM25*- weighted keywords [34].

Our objective through these experiments is twofold: first evaluate the impact of concept-based indexing over keyword-based indexing, and second evaluate the impact of the weighting scheme *cc-idc* on retrieval effectiveness.

## 5.2. Concept-Based Indexing Evaluation

To evaluate the effectiveness of our concept-based indexing disregarding the proposed weighting scheme, we consider our semantic index firstly weighted by *tf*idf* (index *Sem-TF-IDF*) and secondly weighted by Okapi-BM25 (index *Sem-BM25*). Then, we compare the retrieval results from these indexes to those of *Classic-TF-IDF* and *Classic-BM25* baselines respectively. This approach allows isolate concepts contribution from weights contribution.

The evaluation results obtained for these different models are presented in Figure 1.

---

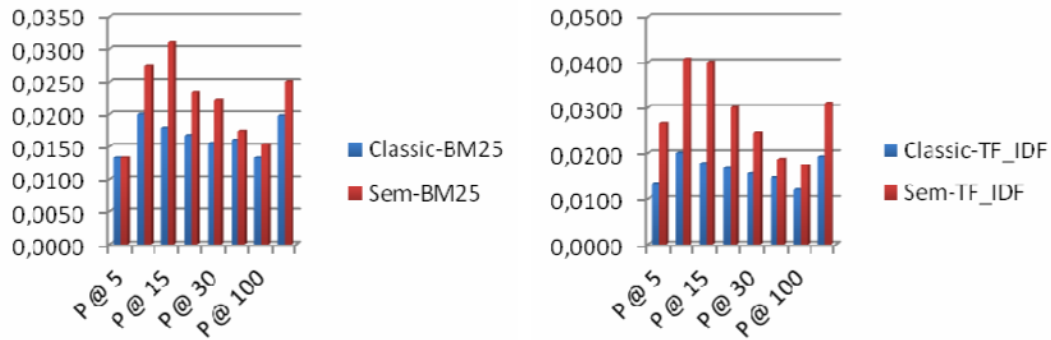

(a) Sem-BM25 vs Classic-BM25          (b) Sem-TF_IDF vs Classic-TF_IDF

Figure 1.  Concept-based vs keyword-based indexing.

According to these results, we noticed that:

- The *Sem-TF-IDF* approach performs better than the *Classic-TF-IDF* baseline. Significant improvement rates (better than 25%) of 100 % for P@5, 103,33 % for P@10, 124 % for P@15, 80% for P@20,  57,27% for P@30, 27,27% for P@50,  44,44% for P@100, and 61,23 % for the MAP are observed.

- Besides, the *Sem-BM25* approach is better than the *Classic-BM25* baseline. Whereas Sem-BM25 and Classic-BM25 performs identically for the P@5 precision, significant improvement rates of 36,67 % for P@10, 75 % for P@15, 40% for P@20, 42,99% for P@30, and 26,8 % for the MAP are observed. Improvement rates for P@50, P@100 although non-significant are positive of 8,33% and 15% respectively.

From these results, it is clear that our semantic indexes (*Sem-TF-IDF* and *Sem-BM25*) are more effective than keyword-based indexes. At this evaluation stage, it is therefore clearly stated that our concept-based indexing approach is more efficient than keyword-based approaches.

Moreover, as shown in Figure 2, the *Sem-TF-IDF* approach performs better than the *Sem-BM25* approach. Improvement rates of 100 % for P5, 48,78 % for P10, 28,55 % for P15, 10% for P@30, 7,69% for P@50, 13,04 for P@100, and 23,60% for the MAP are observed.

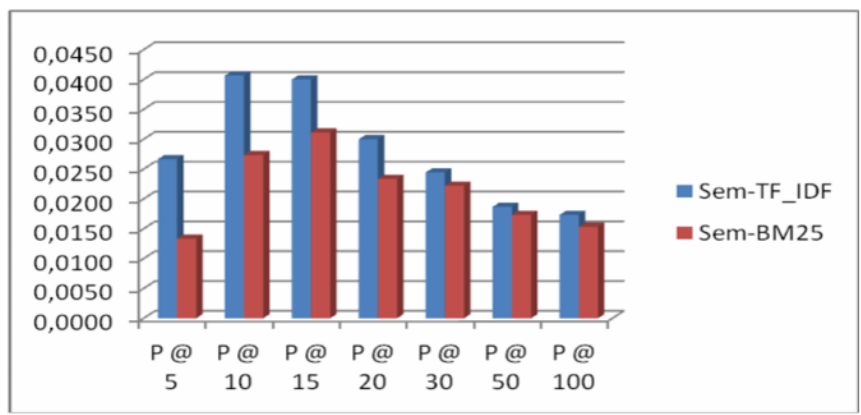

Figure 2. Sem-TF-IDF vs Sem-BM25

132

## 5.3. Concept Weighting Evaluation

The second series of experiments focuses on the evaluation of our concepts weighting scheme introduced in section 4.2.2. Practically, we aim at measuring the impact of the weighting scheme *cc-idc* on the retrieval effectiveness. For this aim, we compare retrieval results from our semantic index weighted by *cc-idc* (*Sem-CC-IDC*) to those of *Sem-TF-IDF* and *Sem-BM25*.

As the proposed weighting scheme *cc-idc* depends on the α weighting parameter, preliminary experiments are necessary to fix the appropriate related value

### 5.3.1. α Selection

We conducted a series of experiments in order to fix the appropriate value for the α weighting parameter. In these experiments, we vary α values between 0 and 1, resulting in different weighting schemes which are successively used to weight our semantic index. The weighted indexes thus obtained are then evaluated through their retrieval results. Evaluation is performed according to the protocol introduced above (section 5.1.2); It is based on two measures: average precision *MP@x* and *MAP,* where *MP@x* is the arithmetic mean of *P@x* ($x$ = 1,2,3,4,5,10,15,20,30,50,100) precisions

Table 5 presents the results of this evaluation. These results show that the overall best retrieval performances (according to *MP@x* and *MAP*) are obtained for α = 0.2.

Table 5. Measuring the impact of α parameter on the retrieval effectiveness.

|         | =0,1   | =0,2   | =0,3   | =0,4   | =0,5   | =0,6   | =0,7   | =0,8   | =0,9   |
|---------|--------|--------|--------|--------|--------|--------|--------|--------|--------|
| **P @ 1**  | **0,1333** | **0,1333** | **0,1333** | **0,1333** | **0,1333** | **0,1333** | **0,1333** | **0,1333** | **0,1333** |
| **P @ 2**  | **0,1333** | **0,1333** | 0,1000 | 0,0667 | 0,1000 | 0,1000 | 0,0667 | 0,0667 | 0,0667 |
| **P @ 3**  | **0,0889** | **0,0889** | **0,0889** | 0,0667 | **0,0889** | **0,0889** | 0,0667 | 0,0444 | 0,0444 |
| **P @ 4**  | **0,0667** | **0,0667** | **0,0667** | 0,0500 | **0,0667** | **0,0667** | **0,0667** | **0,0667** | **0,0667** |
| **P @ 5**  | **0,0800** | **0,0800** | **0,0800** | **0,0800** | **0,0800** | **0,0800** | **0,0800** | 0,0667 | 0,0667 |
| **P @ 10** | 0,0667 | **0,0733** | **0,0733** | **0,0733** | **0,0733** | 0,0667 | 0,0667 | 0,0667 | 0,0467 |
| **P @ 15** | 0,0534 | 0,0534 | 0,0534 | 0,0534 | 0,0534 | **0,0578** | **0,0578** | 0,0534 | 0,0489 |
| **P @ 20** | 0,0400 | 0,0400 | 0,0400 | **0,0433** | **0,0433** | **0,0433** | **0,0433** | **0,0433** | 0,0400 |
| **P @ 30** | **0,0378** | **0,0378** | **0,0378** | **0,0378** | **0,0378** | **0,0378** | 0,0355 | 0,0333 | 0,0333 |
| **P @ 50** | **0,0253** | **0,0253** | **0,0253** | **0,0253** | **0,0253** | **0,0253** | 0,0240 | 0,0240 | **0,0253** |
| **P @**    | 0,0160 | 0,0160 | 0,0160 | 0,0160 | 0,0167 | 0,0173 | 0,0173 | **0,0180** | **0,0180** |
| **MP@x**   | 0,0674 | **0,0680** | 0,0650 | 0,0587 | 0,0653 | 0,0652 | 0,0598 | 0,0560 | 0,0536 |
| **MAP**    | 0,0968 | **0,0987** | 0,0968 | 0,0968 | 0,0970 | 0,0970 | 0,0929 | 0,0861 | 0,0787 |

In the following, we fix the value of α to 0.2. This value favors concept relevance over concept frequency in the corresponding weighting scheme.

### 5.3.2. Experimental Evaluation

The purpose of this step is to evaluate the impact on the retrieval effectiveness of the proposed semantic weighting scheme *cc-idc* over classical weighting schemes *tf*\**idf* and *Okapi-BM25*. For this aim, we compare retrieval results from our semantic index *Sem-CC-IDC* (for α = 0.2) to those of *Sem-TF-IDF* and *Sem-BM25* respectively. The experimental results are represented in Figure 3.

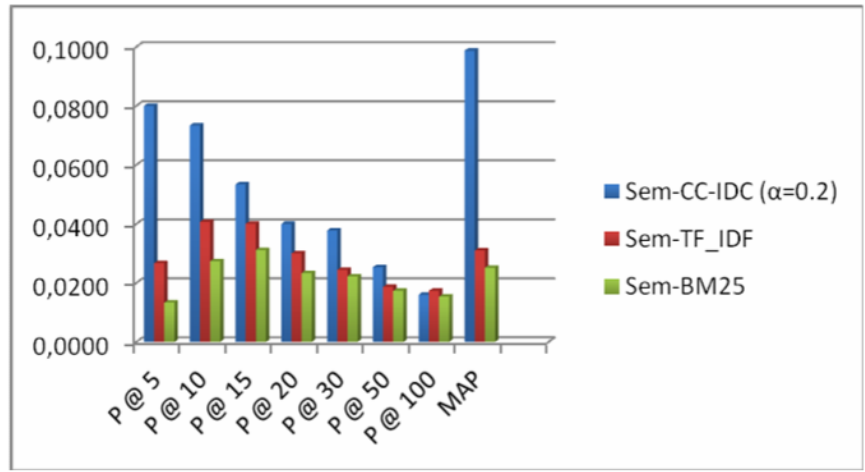

Figure 3. Semantic weighting vs Classical weighting

According to these results, we noticed that:

-   The *Sem-CC-IDC* index performs better than the *Sem-TF-IDF* index. Significant improvement rates (better than 25%) of 200% for P@5, 80,33% for P @10, 33,36% for P@15, 33,33% for P@20, 54.47% for P@30, 35,71% for P@50 and 218,34% for MAP are observed. Nevertheless, a non-significant decreasing rate of 7,69% is obtained at P@100 precision.

-   Moreover, *Sem-CC-IDC* performs better than *Sem-BM25*. Significant performance rates (better than 25%) of 500%, 168.29%, 71.42%, 71.42%, 69.91%, 46,15%, and 293% are observed respectively for P@5, P@10, P@15, P@20, P@ 30, P@50 and MAP. An increasing rate of about 4% is also observed for P@100 precision.

From these results, it is clear that our cc-idc-weighted semantic index (*Sem-CC-IDC*) is more effective than classically-weighted semantic indexes (*Sem-TF-IDF* and *Sem-BM25*). At this evaluation stage, it is therefore clearly stated that our cc-idc-weighting scheme is more effective for concepts than classical *tf\*idf* and *BM25* schemes.

## 6. CONCLUSIONS

In this paper, we have presented a novel approach to automatic concept-based document indexing. Our contribution is twofold: first, we have introduced a novel concept identification approach based on a novel domain-based word sense disambiguation framework that rely on the joint use of WordNet and WordNetDomains, and second we have defined a novel semantic weighting scheme that relies on concept centrality. In our proposal, the centrality of a concept is based on its apparent importance (measured across its frequency of occurrence) in the document on the one hand and on its latent importance (measured across its semantic relatedness to other concepts) in the document on the other hand. Our experimental results showed that the proposed concept-based indexing approach is more effective than classical keyword-based indexing ones; Moreover, our cc-idc-weighting approach (for the fixed value of the weighting parameter ) performs better than classical TF-IDF and BM25 weighting schemes. In future works, we plan first to check to what extent the weighting factor α depends or not on the used document collection, and second to propose a retrieval score for documents that takes into account semantic concepts weights.

## AUTHORS

**Fatiha Boubekeur** is currently assistant professor at the department of Computer Science of Mouloud Mammeri University of Tizi-Ouzou (Algeria). She has received her Ph.D. degree in computer science from University of Toulouse III (France) in 2008. Her research interests are in semantic informatio n indexing and retrieval, text mining, literature-based discovery and user preferences modeling. She has served as a program committee member in RJCRI (CORIA 2008), MEDI 2011, ICWIT 2012, ISPS2013 and as additional reviewer in EGC 2009, FQAS 2009, COSI 2011, COSI 2012, COLING 2012, CORIA 2013.

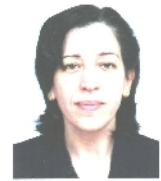

**Wassila Azzoug** is currently a Magister student at the department of Computer Science of Mhamed Bougara University of Boumerdes (Algeria). She has received her Graduate Diploma in Computer Science from the University of Boumerdes in 2008. Her research interests are in semantic information indexing and retrieval.

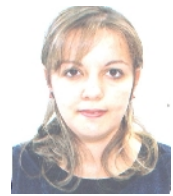